# Breaking of Time Translation Symmetry and Ergodicity, and Entropy decrease in a Continuous Time Crystal Driven by Nonreciprocal Optical Forces

Tongjun Liu[1], Venugopal Raskatla[1], Jinxiang Li[2],
Kevin F. MacDonald[1], Nikolay I. Zheludev[1, 2]

[1]Optoelectronics Research Centre and Centre for Photonic Metamaterials, University of Southampton, Highfield, Southampton, SO17 1BJ, UK

[2]School of Physical and Mathematical Sciences, Nanyang Technological University, Singapore, 637378, Singapore

e-mail: zheludev@soton.ac.uk

Nonreciprocal nonequilibrium process are attracting growing interest in sociology, animal behaviour, chemistry, and nanotechnology, and may have played a role in the origin of life. It is less widely recognized, however, that in open systems light can induce nonreciprocal predator-prey like forces between nanoparticles. Such forces provide access to the continuous time crystal state of matter, which has been demonstrated in a plasmonic metamaterial array of nanowires wherein light triggers a spontaneous 'mobilization' transition to the robust oscillatory state, breaking time translation symmetry [1]. Here, we report on the first experimental study of the transient dynamics of light-induced mobilization and demobilization in a time crystal. By analysing time resolved phase trajectories of the system of nanowires, we show that the mobilization transition is accompanied by breaking of continuous time translation symmetry and ergodicity, and a decrease in the entropy of motion. This insight into the transient dynamics of a nonreciprocity-driven time crystal is relevant to optical "timetronics" – an information and communications technology paradigm relying on the unique functionalities of time crystals, and applications of the interacting nanowire oscillator platform to modelling a wide range of nonreciprocal processes from many-body dynamics to the early stages of matter-to-life transitions.

**Introduction**

Nonreciprocal processes in systems out of equilibrium are attracting growing interest across the entire domain of scientific research from sociology to nanotechnology, including a recent suggestion that nonreciprocal interactions may have played a critical role in the origin of life, i.e. the matter-to-life transition. Nonreciprocal interactions are universally present in living systems and underpin the behaviour of inhomogeneous social groups, swarming, and hierarchical and predator-prey dynamics. Nonreciprocal interactions arise naturally in molecular systems, for example due to diffusion and kinetic asymmetries as in the 'predator–prey chasing' among oil droplets of different chemistries, and they lead to time-dependent phases as illustrated by the dynamics of interacting autonomous vehicles. In many-body ensembles, nonreciprocal interactions initiate symmetry breaking and lead to pattern formation, hysteresis, and phase transitions [2-5]. In open systems, light can induce nonreciprocal predator-prey-like forces between nanoparticles due to differential scattering [6-8], and it has recently been demonstrated that such interactions facilitate realization of the continuous time crystal state of matter in a plasmonic nanowire metamaterial [1, 9].

A time crystal, first discussed by Frank Wilczek in 2012, is a many-body interacting system that exhibits a spontaneous 'mobilization' transition to a robust oscillatory state, breaking time translation symmetry under an arbitrarily small change in the external driving force. Two types of time crystals are defined: "Discrete" time crystals break discrete time-translation symmetry by oscillating at a sub-harmonic frequency of periodic external driving force; "Continuous" time crystals break continuous time-translation symmetry by transitioning from disorder to robust periodic motion (with arbitrary phase) in response to a time-invariant force.

Various realizations of time crystals using spins, Bose-Einstein condensates, magnons, the qubits of a quantum computer, and other exotic systems have been demonstrated [10-13]. However, the continuous time crystal realized in Ref. [1] is uniquely driven by nonreciprocal, predator-prey-like, optical forces in a metamaterial array, which arise between two types of nanowire oscillator alternately decorated with dissimilar plasmonic nanoparticles. Above a threshold value of light intensity, the nonreciprocal forces trigger a spontaneous transition in the nanowire ensemble, from a state of uncorrelated stochastic thermal motion to the 'mobilized' state of persistent, synchronized, high amplitude periodic oscillation [9].

However, the transient dynamics of light-induced mobilization and demobilization in this nonreciprocity-driven continuous time crystal have not been studied. Here, by monitoring transitions in the nanowire ensemble in the time domain, we experimentally obtain the mobilization and demobilization times of the time crystal state. We provide the first experimental study of ergodicity-breaking in a time crystal by mapping the phase of the demobilized, mobilized and intermediate transient states. Finally, we demonstrate that the entropy of the nanowires' motion is locally reduced upon mobilization. We show that experimental observations are extremely well described by the Langevin model for a thermally driven and nonreciprocally coupled ensemble of linear oscillators.

Intriguingly, the above characteristics of the time crystal transition – reduction of entropy, and spontaneous breakage of time translation symmetry and ergodicity, are also key characteristics of living matter.

**Experimental study of mobilization and demobilization dynamics**

We studied a metamaterial array comprising alternately dissimilar nanowires decorated with plasmonic nanorods, fabricated by focused ion beam milling from a gold-coated silicon nitride membrane. The nanorods are arranged to form split Π-shaped metamolecules each supported on a pair of neighbouring nanowires, as illustrated in Fig. 1. The nanowires' mutual out-of-plane displacement reconfigures the metamolecules, strongly affecting their plasmonic resonances and thereby the optical properties of the array. The dimensions of the metamolecule are specified such that light at a wavelength of $\lambda = 1.55$ µm induces nonreciprocal out-of-plane forces between its two parts located on neighbouring nanowires. Upon illumination, the nanowire supporting a single gold nanorod (of the three in each metamolecule) becomes – in terminology of nonreciprocal dynamics [2] – a "predator" chasing the "prey" – i.e. the nanowire supporting the other (two gold nanorod) part of the metamolecule, which is forced to move away from the "predator". The strength of this nonreciprocal interaction increases with light intensity and upon reaching a threshold value triggers spontaneous synchronization of nanowire (oscillator) movement across the array [9].

Below the threshold intensity, the nanowires exhibit independent thermomechanical stochastic movement resulting from the momentum transfer of flexural phonon creation and annihilation. According to the Langevin theory of thermally driven oscillators, these thermal fluctuations are most profound at the natural frequency of the oscillator. In the present metamaterial array the resonance frequencies $\omega_{0i}$ of the individual nanowires' out-of-plane motion lie between $2\pi \times 852$ kHz and $2\pi \times 858$ kHz – a dispersion due to fabrication imperfection and inhomogeneity in the membrane's intrinsic stress. At room temperature the oscillators move independently with time-averaged resonant amplitudes $\langle x_T \rangle = \sqrt{\frac{k_B T}{m_i \omega_{0i}^2}}$ of about 300 pm, where $m_i$ is the nanowire effective mass, $T$ is temperature, and $k_B$ is the Boltzmann constant. These picometric fluctuations measurably change the reflectivity and transmissivity of the array [14], which depends sensitively on the mutual positions of the nanowires, providing a means for detecting the state of array optically: upon mobilization, the nanowires begin move in unison, their amplitudes of oscillation increases and the frequency spectrum of their motion collapses.

The metamaterial was mounted in a low-vacuum cell with pressure maintained at ~$10^{-4}$ mbar to suppress air damping of nanowire movement. It was illuminated at normal incidence with laser light (polarized parallel to the nanowires and focused to a ~5 μm spot) at a wavelength $\lambda = 1.55$ µm, close to the plasmonic resonance of the metamolecules. The transmissivity $T$ of the array was monitored by an InGaAs photodiode with 125 MHz bandwidth.

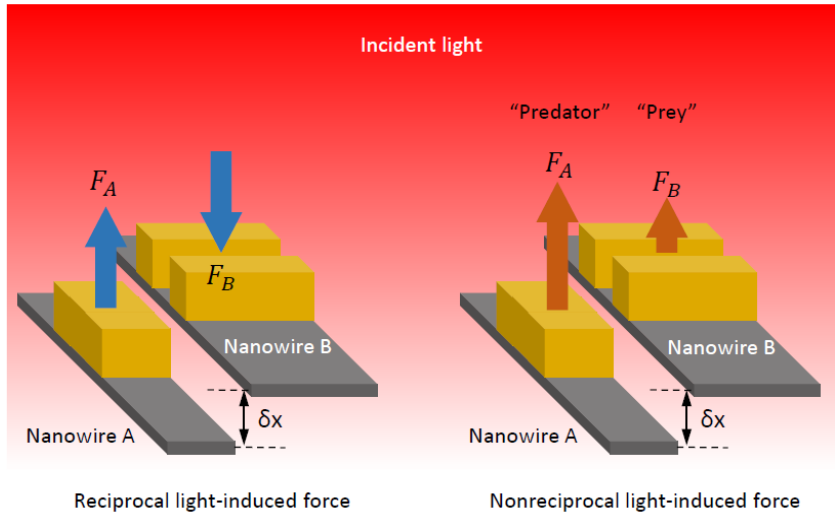

**Fig. 1: Predator-Prey dynamics induced by light in a nanowire plasmonic metamaterial.** When stochastic thermal motion of the nanowires results in a mutual out-of-plane displacement $\delta x$ between them, light scattered on the supported plasmonic nanorods asymmetrically changes the balance of optical forces acting on nanowires. If the additional forces are reciprocal $\mathbf{F}_A = -\mathbf{F}_B$, i.e. equal in magnitude and opposite in direction (left panel), the motion of the nanowires remains unsynchronized. If the additional forces are nonreciprocal (asymmetric) $\mathbf{F}_A \neq -\mathbf{F}_B$, a predator-prey dynamic is established, and above the threshold light intensity nanowire oscillations synchronize.

With adiabatically increasing incident light intensity the array exhibits a spontaneous mobilization phase transition to the synchronized oscillatory state at a threshold incident power $P = 39$ µW. With adiabatically decreasing power, the demobilization transition occurs at $P = 36$ µW.

In what follows, we explore the dynamics of these transitions by applying sharp "square wave" step changes in incident laser power across the thresholds (as shown by the blue traces in Figs. 2a-c), using an electro-optic modulator. We record time series data of the photodiode current and then filter the signal to retain only frequency components in the range $\Delta f$ from 850 to 870 kHz, encompassing the natural resonance frequencies of all illuminated nanowires' out-of-plane motion. By normalizing this signal against the DC component of the photodiode current, which is proportional to sample transmissivity $T_0$ at the fixed average laser power $P = 29.4$µW, we arrive at normalized transmissivity change $\delta T(t)/T_0$ – a parameter that depends directly on the state of nanowire array, $\delta T(t)$ being transmissivity change in the spectral range $\Delta f$ associated with the out-of-plane oscillation of the nanowires.

Below the $P = 39$ µW synchronization threshold power, the monitored normalized transmissivity change fluctuates with low amplitude about zero due to the incoherent thermal motion of the nanowires and detection system noise, as shown by the orange trace in Fig. 2a (where incident laser power is switched between sub-threshold levels of 21 and 37 µW).

Figure 2b shows transmissivity dynamics at the onset of synchronization, when incident laser power is increased abruptly from ~17 to ~40 µW, which is to say from markedly below to just above threshold. While the power is above threshold, the amplitude of transmissivity modulation slowly increases (mobilization). The rise time is unstable from one synchronization cycle to the next, varying from ~20 to ~35 ms (from ~17,000 to ~30,000 oscillations cycles) and the amplitude of oscillation in the synchronized (mobilized) state is unstable. When the incident peak power is further above threshold (~43 µW in Fig. 2c) the synchronization dynamics stabilize – the mobilization transition occurs faster, with a rise time reduced to 15 ms. In contrast, the decay/relaxation time of the reverse (demobilization) transition from the synchronized state to the incoherent (thermally fluctuating) state, does not depend on incident laser power.

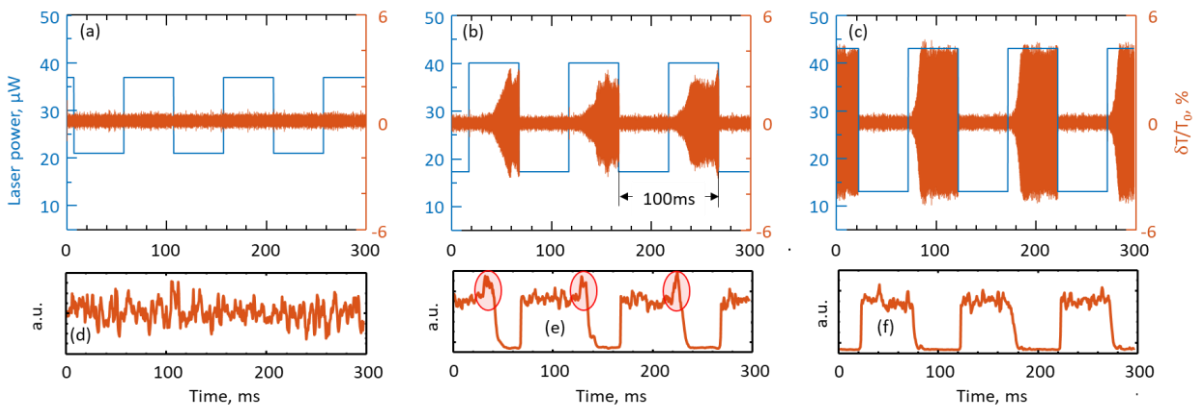

**Fig. 2: Dynamics of optically-induced spontaneous mobilization and demobilization transitions in the metamaterial continuous time crystal.** (a-c) Relative transmissivity change $\delta T(t)/T_0$ of the metamaterial array [orange lines, right-hand axis] related to the out-of-plane motion of nanowires as a function of time while incident laser power is switched between different hight and low levels [blue lines, left-hand axis]. (d-f) Corresponding dynamics of transmissivity spectral entropy.

In Figs. 2d-f, we plot the instantaneous spectral entropy of transmissivity modulation which is a measure of the disorder of oscillations, or randomness, of the system. It is defined as: $H = -\sum_{m=1}^{N} P(f_m) \log_2 P(f_m)$, where the probability distribution $P(f_m) = \frac{S(f_m)}{\sum_i S(f_i)}$, and $S(f_m)$ is the power spectrum of the process. The entropy decreases slowly from the onset of mobilization and increases rapidly at the demobilization transition. These data show that the mobilized state is characterized by a stable entropy level below that of the disordered state. We note here that when the peak incident power is just above the synchronization threshold (Fig. 2b), entropy initially increases above the disordered state level (the spikes highlighted in Fig. 2e) before falling to the lower mobilized state level. This behaviour indicates a highly nonequilibrium regime of motion at the onset of mobilization.

With an abrupt decrease in incident laser power from above to markedly below the mobilization threshold, the oscillators revert to the regime of unsynchronized, incoherent thermomechanical fluctuation. The amplitude of oscillation decays exponentially with a relaxation, or demobilization, time $1/\gamma$, observed to be ~60 µs (~50 cycles, Fig. 3a). The oscillators de-phase at a rate of $\pi/(2\delta\omega_0)$, where $2\delta\omega_0$ is the inhomogeneous broadening of the array, which is significantly shorter than the relaxation time $1/\gamma$, leading to a complex demobilization process.

To further reveal the dynamics of mobilization, we plot transmissivity phase diagrams ($\delta T$ against $\partial\delta T/\partial t$) for different stages of the transition (Fig. 3c-e). In the disordered state, i.e. while incident laser power is below the mobilization threshold, characteristically stochastic motion is observed with a Gaussian phase distribution (Fig. 3c). When incident power is increased above threshold, there is a transition (Fig. 3d) to a limit cycle of stable synchronized motion (mobilized state, Fig 3e), reflecting a loss of ergodicity from a state in which many individual modes of oscillation are simultaneously present to one of periodic harmonic motion.

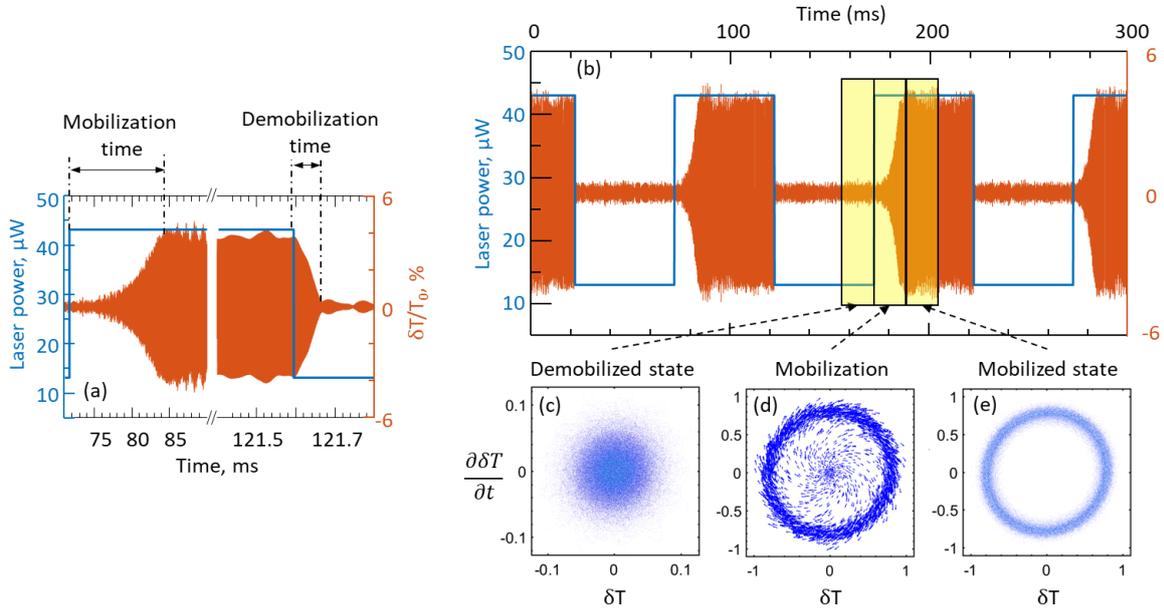

**Fig 3: Transient mobilization and demobilization times.** (a, b) Transmissivity oscillation dynamics during mobilization and demobilization. Plates (c-e) show transmissivity phase diagrams: (c) of the disordered state; (d) during mobilization; (d) the mobilized time crystal state. Phase diagrams are derived from the analysis of the segments indicated by the yellow shaded boxes in (b) and formed from hundreds of repeated mobilization cycles.

Importantly, the results presented in Fig. 3 also demonstrate that continuous time translation symmetry is broken by the mobilization transition, through the randomness of the phase of the mobilised state. Indeed, Fig. 3d illustrates that the system "winds-up" to the mobilized state through a different spiral phase 'trajectory' in each cycle with an arbitrary initial phase prescribed by thermal fluctuations, while the distribution of points in Fig. 3e illustrates the random phase of oscillation at a fixed interval after incident laser intensity exceeds the mobilization threshold.

**Modelling the transient dynamics of mobilization and demobilization**

The main features of the metamaterial time crystal's mobilization and demobilization dynamics can be understood from a model describing them as a nonreciprocal phase transition induced by nonconservative radiation pressure forces among the plasmonic metamolecules. For the modelling we consider a set of 24 coupled oscillators. Thermal motion at non-zero temperature is accounted for by assuming that the oscillators are connected to a common bath (in experiment, the frame of the nanowire array) at temperature $T$. Such a system is described by the Langevin model for linear oscillators with frequencies $\omega_{0i}$, masses $m_i$ and loss parameters $\gamma_i = \omega_{0i}/Q_i$:

$$\ddot{x}_i + \gamma_i \dot{x}_i + \omega_{0i}^2 x_i + \sum \xi_{ij}(x_i - x_j) = \sqrt{\frac{2k_B T \gamma_i}{m_i}} \eta_i(t) \qquad (1)$$

where $Q_i$ and $\eta_i(t)$ are respectively oscillator quality factors and normalized white noise terms, and the parameter $\xi_{ij}$ describes light-induced coupling between oscillators.

We analyse the behaviour of this system under various coupling conditions by numerically solving Eq. (1), using oscillator parameters $m_i = 1$ pg, and $\gamma_i = \omega_c/Q$; $\omega_c = 2\pi \times 860$ kHz, $Q = 1000$, close to those of the experimental system of Ref. [1]. In Maxwell stress tensor calculations [9], we have evaluated the nonreciprocal electromagnetic forces acting between gold nanorods and the concomitant nanowire coupling terms $\xi_{ij}$. To account for inhomogeneous broadening of the oscillator ensemble we assume frequencies that are randomly distributed in the interval from 855 to 867 kHz about the central frequency $\omega_c = 2\pi \times 860$ kHz. We also assume that each oscillator is coupled only to its near neighbours. The results of this modelling are shown in Fig. 4, in terms of the collective motion amplitude $X(t) = \sum_{i=1}^{i=12}(x_{2i}(t) - x_{2i-1}(t))$ as related to the experimentally measured time dependent normalized transmissivity $\delta T(t)/T_0$, and spectral entropy.

Our modelling results collaborate well with experimental data. Considered together, they demonstrate that nonreciprocal coupling is the main mechanism of mobilization in the metamaterial time crystal, whereby the onset of synchronization does not require nonlinearity in the system. Nonlinearity only manifests here when oscillation amplitude is much higher that the thermal motion amplitude, serving to stabilize the amplitude of motion in the mobilized state.

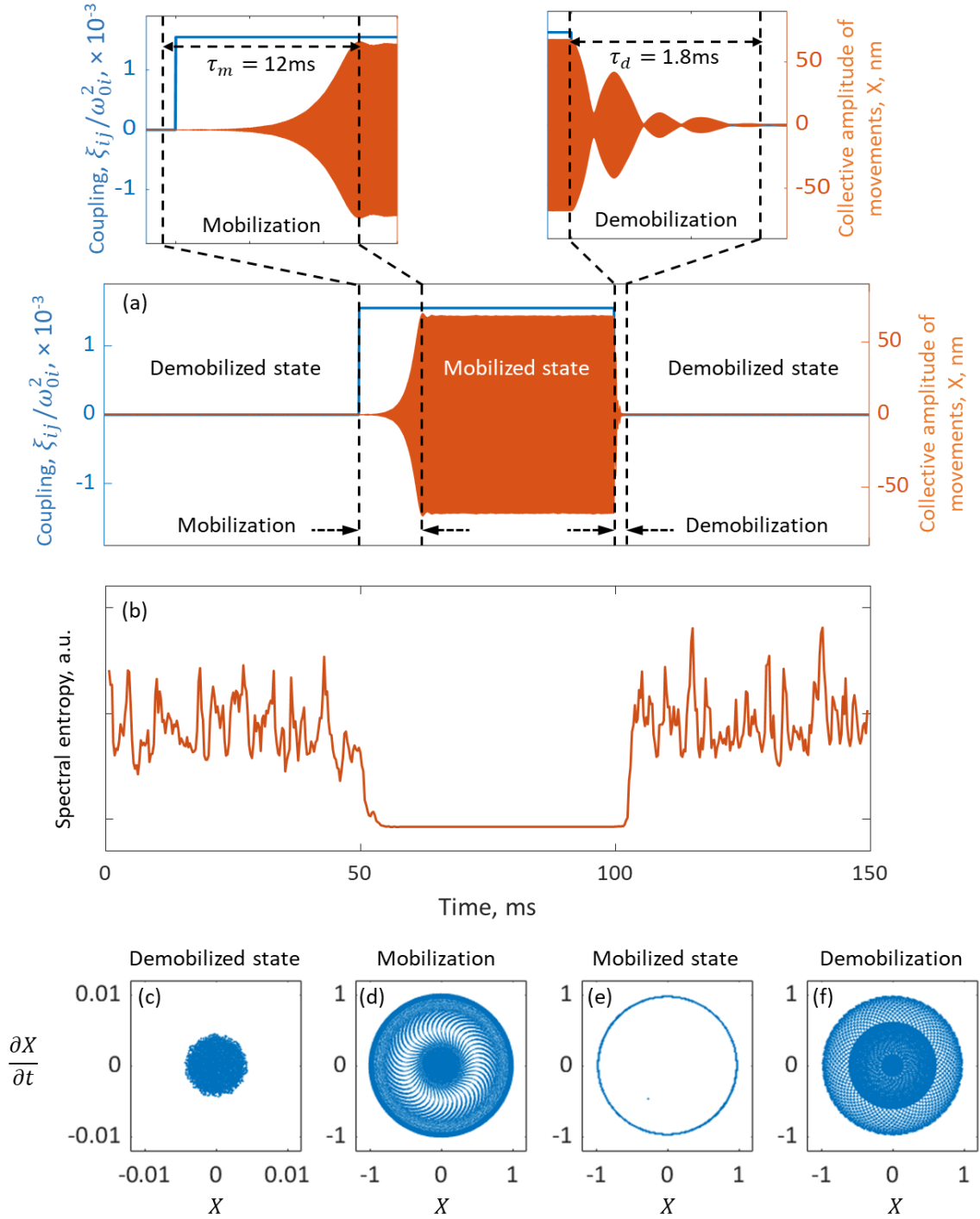

**Fig. 4: Dynamics of an ensample of 24 thermally driven, nonreciprocally coupled Langevin oscillators.** Time dependence of (a) collective motion amplitude *X(t)* [orange traces] and (b) spectral entropy of collective movement, as coupling strength [blue traces in (a)] is cycled through step-changes from zero to a value above the array's synchronization threshold, and back. (c-f) Collective motion phase diagrams: (f) for the demobilized state; (g) during mobilization; (h) for the mobilized state; (h) during demobilization.

**Conclusion**

We present the first study of the dynamics of mobilization and demobilization transitions between the states of incoherent thermal motion and persistent synchronized oscillation in a nano-opto-mechanical metamaterial time crystal - an array of alternately dissimilar nanowires with variable, optically-controlled nonreciprocal coupling between them.

By analysing the time-domain response of the system, we have shown experimentally that the speed of mobilization, from incoherent thermomechanical fluctuation to synchronized oscillation of the array, increases with incident light intensity (Fig. 2b, c); and that the mobilization transition is slower than the intensity-independent demobilization transition to the disordered state, which occurs upon withdrawal of illumination and which is controlled by damping and dephasing of the inhomogeneous broadened nanowire oscillator ensemble. By resolving the phase of the system, we have shown that the mobilization transition is accompanied by ergodicity breaking whereby, with increased optical coupling, the oscillation frequencies of individual nanowires converge to a single-phase state.

The remarkable correlation between experiments and a computational model for oscillator ensemble synchronization, in which coupling coefficients are derived from Maxwell stress tensor simulations of the optical forces between plasmonic nanorods as in the experimental system, confirms that the mobilization transition depends upon nonreciprocal interactions underpinned by asymmetric light scattering, and does not require nonlinearity in the system to kick-start mobilization.

We note that this mechanism of realizing a continuous time crystal requires the presence of two dissimilar components – in experiment, nanowires supporting different parts of the plasmonic metamolecules – with predator-prey dynamics that are characteristic of a wide variety of systems in which nonreciprocity-driven symmetry breaking, pattern formation, hysteresis and phase transitions are observed, from inhomogeneous social group and swarms of animals to interacting autonomous vehicles and nonequilibrium chemical processes. Interestingly, the observed decrease in the entropy of the optically mobilized time crystal state may be compared with the entropy dynamics of plants growing under the influence of the sun, wherein disordered matter evolves into highly organized living matter [15], illustrating the recently proposed view [5] that nonreciprocal interactions may have played a critical role in the matter-to-life transition.

The realization of optically controlled time crystal matter in the form of nanowire metamaterials offers new opportunities for developing the domain of optical "timetronics" as a new information and communications technology paradigm relying on the unique functionalities of time crystals. The interacting nanowire oscillator platform also opens a path to experimental exploration and modelling of a wide range of nonreciprocal processes, from many-body dynamics to the early stages of matter-to-life transitions.

This work was supported by the UK Engineering and Physical Sciences Research Council (grant EP/T02643X/1 – NIZ, KFM) and the National Research Foundation Singapore (grant NRF-CRP23-2019-0006 – NIZ).